\documentclass[aps,prl,reprint,superscriptaddress,showpacs]{revtex4-1}

\usepackage{graphicx} 
\usepackage{dcolumn}
\usepackage{amsfonts}
\usepackage{amssymb}
\usepackage{amsmath}
\usepackage{bm}
\usepackage{epsf}
\usepackage{epstopdf}
\newcommand{\be}{\begin{equation}}
\newcommand{\ee}{\end{equation}}
\newcommand{\bea}{\begin{eqnarray}}
\newcommand{\eea}{\end{eqnarray}}

\begin{document} 

\title{A Strongly Dipolar Bose-Einstein Condensate of Dysprosium}

\author{Mingwu Lu}
\affiliation{Department of Applied Physics, Stanford University, Stanford CA 94305}
\affiliation{E. L. Ginzton Laboratory, Stanford University, Stanford CA 94305}
\affiliation{Department of Physics, University of Illinois at Urbana-Champaign, Urbana, IL 61801}

\author{Nathaniel Q. Burdick}
\affiliation{Department of Applied Physics, Stanford University, Stanford CA 94305}
\affiliation{E. L. Ginzton Laboratory, Stanford University, Stanford CA 94305}
\affiliation{Department of Physics, University of Illinois at Urbana-Champaign, Urbana, IL 61801}

\author{Seo Ho Youn}
\affiliation{Department of Physics, University of Illinois at Urbana-Champaign, Urbana, IL 61801}
\author{Benjamin L. Lev}
\affiliation{Department of Applied Physics, Stanford University, Stanford CA 94305}
\affiliation{E. L. Ginzton Laboratory, Stanford University, Stanford CA 94305}
\affiliation{Department of Physics, University of Illinois at Urbana-Champaign, Urbana, IL 61801}
\affiliation{Department of Physics, Stanford University, Stanford CA 94305}

\begin{abstract}
We report the Bose-Einstein condensation (BEC) of the most magnetic element, dysprosium.  The Dy BEC is the first for an open $f$-shell lanthanide (rare-earth) element and is produced via forced evaporation in a crossed optical dipole trap loaded by an unusual, blue-detuned and spin-polarized narrow-line magneto-optical trap.  Nearly pure condensates of 1.5$\times$$10^{4}$ $^{164}$Dy atoms form below $T = 30$~nK.  We observe that stable BEC formation depends on the relative angle of a small polarizing magnetic field to the axis of the oblate trap, a property of trapped condensates only expected in the strongly dipolar regime.  This regime was heretofore only attainable in Cr BECs via a Feshbach resonance accessed at high magnetic field.  

\end{abstract}
\date{\today}
\maketitle

The interplay between emergent electronic or spatial crystallinity, magnetism, and superfluidity is central to some of the most interesting materials of late, e.g., cuprate and iron-based superconductors, strontium ruthenates, and solid helium~\cite{Fradkin2009,*Fradkin2010,*Chan04,*Davis09}. Quantum degenerate gases possessing strong dipole-dipole interactions (DDI)~\cite{PfauReview09} are thought to provide access to strongly correlated quantum phases involving quantum magnetism, spontaneous spatial symmetry breaking, and exotic superfluidity.  Supersolid~\cite{Pupillo10,*Pollet10,*Duan2010,*Hoffsetter11} and quantum liquid crystal~\cite{Fregoso:2009,*Quintanilla:2009,*Liu10,*DasSarma10c} phases may be accessible using strongly dipolar Bose and Fermi gases; with the realization of such quantum gases, we may shed light on characteristics of these phases difficult to observe in their condensed matter settings~\cite{Chan04,*Davis09,*Fradkin2009}.

We report the first realization of a strongly dipolar quantum gas at low field, a Bose-Einstein condensate (BEC) of $^{164}$Dy, which is an atom with unsurpassed dipole moment $d = 10$$\mu_{B}$, where $\mu_{B}$ is the Bohr magneton~\footnote{Fermionic Dy's magnetic moment is the largest of all elements, and bosonic Dy's magnetic moment is equal only to Tb's~\cite{Martin:1978}.}.  For comparison, Rb and Cr's moment equals 1$\mu_{B}$ and 6$\mu_{B}$, respectively.  The Dy BEC is the first quantum gas of a highly complex, open $f$-shell lanthanide, and opens a new frontier for exploring scattering behavior involving DDI-induced universality~\cite{Ticknor08,*Bohn09,*Ticknor10,*Greene11b}, electrostatic anisotropy~\cite{Newman2011,Kotochigova11}, and an open $f$-shell submerged under closed $s$-shells.  These complex atomic properties evidently do not prevent Bose-condensation, and exploring such complex collisional physics will greatly aid  attempts to understand collisional behaviors of ultracold polar molecules, whose Bose-condensation has yet to be achieved.

Unlike the $^{52}$Cr dipolar BEC~\cite{Pfau05CrBEC}, we show that the $^{164}$Dy BEC reaches the strongly dipolar regime $\epsilon_{dd}$$=$$\mu_{0}\mu^{2}m/12\pi\hbar^{2}a_{s}$$>$1 without careful minimization of the $s$-wave scattering length $a_{s}$ using a high-magnetic field Feshbach resonance~\footnote{For a zero-field (as yet unmeasured) $a_{s}$ equal to $^{52}$Cr's $a_{s}\approx100a_{0}$, $\epsilon^{\text{Dy}}_{dd}=1.3$ while $\epsilon^{\text{Cr}}_{dd}=0.15$.  ($a_{0}$ is the Bohr radius.)  $\epsilon^{\text{Cr}}_{dd}\approx1$ has been achieved using a 589 G Feshbach resonance~\cite{Pfau07CrBECferrofluid}.}.  The use of Feshbach resonances can lead to detrimental three-body loss~\cite{Pfau07CrBECferrofluid} and precludes the study of ultracold dipolar physics near zero field. Moreover, ultracold dysprosium  suffers no chemical reactions like certain polar molecules~\cite{Ospelkaus2010}.

Short-wavelength optical lattices confining Dy are capable of exploring uncharted strongly correlated phases beyond the familiar Mott insulator at half-filling.  Namely, density waves of various filling factors and lattice supersolids may now be accessible~\cite{PfauReview09,Pupillo10,Pollet10,Duan2010,Buchler11} without multilayer lattice enhancement~\cite{Pfau11}.   Moreover, creation of unconventional and anisotropic superfluids~\cite{Lewenstein09,*Bonitz10,*Ticknor11} as well as explorations of 1D strongly correlated gases~\cite{Sondhi09,*Zoller2010},  trap instabilities~\cite{PfauReview09}, spin textures and topological defects~\cite{Pi09,*Santos09,*Machida10,*Jezek10}, roton modes~\cite{Bohn07,*Bohn08}, and emergent structure in layered dipolar gases~\cite{Wunner10,*Bohn11,*Blume11} are among the phenomena now within the realm of experimental possibility.

\begin{figure*}[t]
\includegraphics[width=1\textwidth]{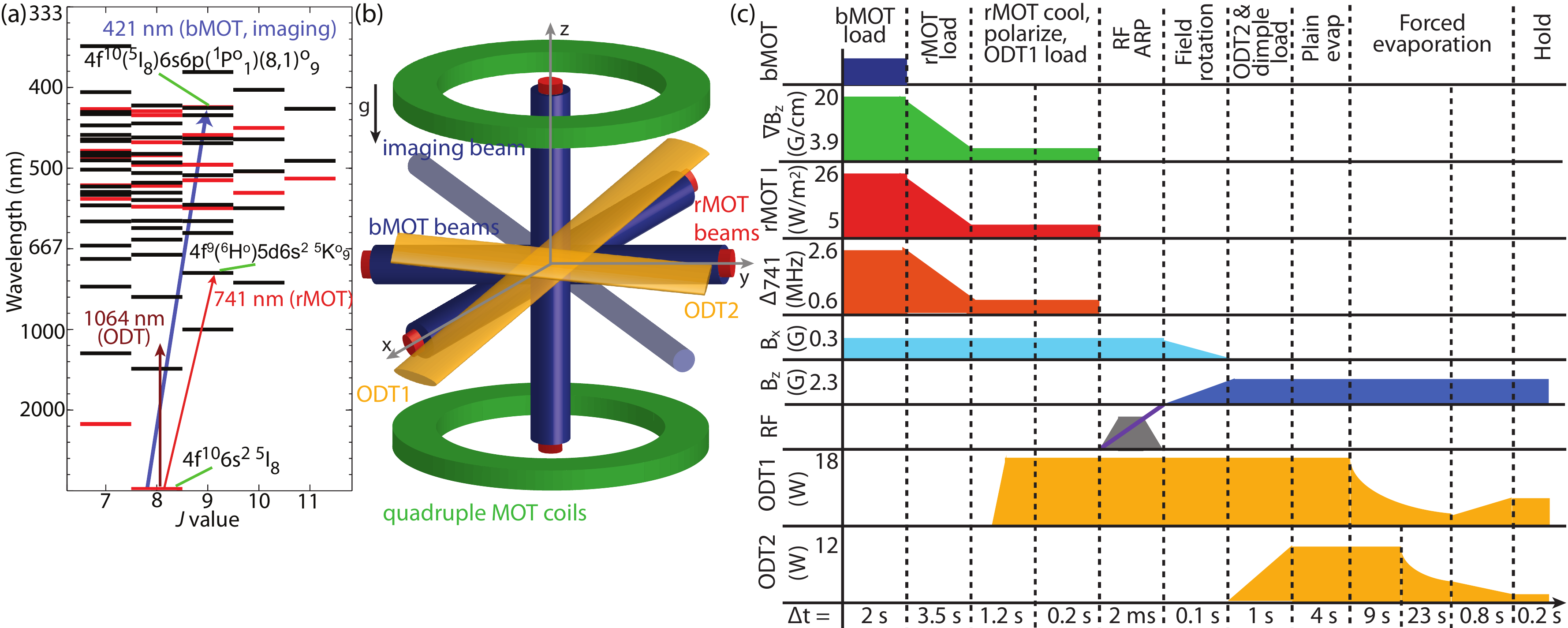}
\caption{(color online). (a) The relevant $^{164}$Dy level scheme.  The nuclear spin is $I = 0$ and the total electronic angular momentum is $J=L+S=8$ with $L = 6$ and $S = 2$.  Experimental setup (b) and typical timing diagram (c). The trapezoid and diagonal line in the RF row represent the RF power ramp and the 0.7 MHz sweep about 5.65 MHz, respectively.}
\label{Scheme}
\end{figure*}

The cooling procedure follows aspects of the Sr, Ca, and Yb BEC experiments~\cite{YbBEC07b,*CaBEC09,*SrBECKillian09,*SrBECSchreck09}, and we only provide essential experimental details before briefly discussing some  of the $^{164}$Dy BEC properties.  The BEC of $^{164}$Dy, natural abundance 28.2\%, is formed by forced evaporative cooling in a crossed optical dipole trap (cODT).  The cODT is loaded with ultracold atoms from an unusual, blue-detuned narrow-line magneto-optical trap (MOT)~\cite{Berglund:2008} formed on the 1.8-kHz 741-nm line~\cite{Lu2011}, the demonstration of which for Dy we report  for the first time.   This red-wavelength MOT (rMOT) is loaded by a repumperless blue-wavelength MOT (bMOT) formed on the broad, 421-nm transition.  A Zeeman slower and transverse cooling stage, both at 421~nm, slow and collimate an atomic beam from a 1250~$^{\circ}$C oven.  Details  regarding the bMOT are provided in Refs.~\cite{Lu2010,*Youn2010a}, and a subsequent publication will thoroughly describe the Dy rMOT and cODT.  

Figure~\ref{Scheme} depicts the relevant Dy energy level diagram with cooling and trapping lasers as well as a timing diagram for the experiment. The bMOT captures 5$\times$10$^{8}$ $^{164}$Dy atoms, 70\% of which, in steady-state, are $\sim$1~mK metastable atoms concurrently confined within the bMOT's magnetic quadrupole trap.  We operate the bMOT at 1.4$\times$10$^{8}$ total atoms for this work.  Of these, 7$\times$10$^{7}$  are captured in the rMOT---whose beams are overlapped with the bMOT's---during the loading phase lasting 3.5 s.  Concurrently,  metastable Dy  decays to the ground state with a  time constant of 2.3 s$^{-1}$~\cite{Lu2010}.  

The rMOT differs from most narrow-line MOTs, e.g., Ref.~\cite{YeSr03}, in that the cooling laser is blue-detuned from resonance.  For heavy, highly magnetic atoms possessing a narrow optical transition, a spin-polarized MOT may form below the center of the quadrupole field such that gravitational, magnetic, and optical forces  balance each other.  This technique, first employed for an 8-kHz-wide line in Er ($\mu = 7\mu_{B}$)~\cite{Berglund:2008}, creates an ultracold, dense gas at a position determined by the laser detuning such that the magnetic field  Zeeman-shifts the otherwise red-detuned maximally weak-field-seeking state (MWS) to the blue of the laser frequency.  The cloud shifts position in the magnetic gradient as a function of laser detuning, but the laser linewidth need not be as narrow as the transition, as often the case for narrow-line MOTs.

The Er experiment captured 10-20\% of the $10^{5}$ atoms loaded into the Er bMOT~\footnote{A.~J.~Berglund, private communication (2009).}  by fixing the rMOT laser intensity and detuning while ramping down the quadrupole field before recompression.  We improve the capture efficiency of this technique by sweeping the laser's blue-detuning from $\Delta_{\text{741}}=+2.6$ MHz to $+0.6$~MHz while reducing the laser intensity (quadrupole gradient) from $I=26$ W/m$^{2}$ ($\nabla B_{z}=20$ G/cm) to 4.8 W/m$^{2}$ (3.9 G/cm)~\footnote{The bMOT  is unaffected by the rMOT lasers.}.   We capture 50\% of the total bMOT-cooled population by following this procedure.   We hold this configuration for 1.2 s to cool to 12~$\mu$K and spin-purify the gas in $m_{J}=8$, as determined by subsequent Stern-Gerlach (SG) measurements.  While ultracold, the gas does not reach the line's 84-nK Doppler limit; similarly, the Er rMOT's  2 $\mu$K  temperature was substantially larger than its 190-nK Doppler limit~\cite{Berglund:2008}.  

Reference~\cite{Berglund:2008} suggests that the seemingly hotter temperature arises due to momentum kicks from imperfectly extinguishing  the quadrupole field before time-of-flight (TOF) measurements.  However, these hotter temperatures may also arise from dipolar relaxation heating.  Indeed, the plain evaporation of Dy after optical dipole trap loading implies the gas is no colder than 10 $\mu$K when in the rMOT.   Inelastic dipolar relaxation rates scale strongly with dipole moment and are expected to be quite rapid for Dy and Er~\footnote{Indeed, we observe in a 0.3 G field a 25-fold reduction of ODT trap lifetime in the MWS versus the maximally strong-field seeking state.}.   We speculate that the gas heats by releasing Zeeman energy as the MWS state relaxes to lower-energy Zeeman states, while the rMOT lasers re-cool and repump to the MWS state.  An equilibrium in temperature and spin polarization is reached for a given $\nabla B_{z}$, $\Delta_{741}$, and $I$, and future numerical simulations will investigate this process~\footnote{A. J. Berglund and B. L. Lev, in preparation.}.

We employ two horizontal ODTs, ODT1 and ODT2, to form the cODT at a position beyond the edge of the rMOT.  Intersecting the ODTs with the rMOT during loading repels the gas due to a $\sim$10-kHz Stark shift of the 741-nm line, which repositions the atoms in the quadrupole field to compensate the change in effective rMOT detuning.  The optimal loading strategy thus involves setting the 18 W ODT1's propagation height roughly $r=1$ mm below the rMOT center and fine-tuning the transfer efficiency with the 741-nm laser frequency while preserving spin-polarization and minimizing gas temperature.   

The  transfer efficiency from the rMOT to ODT1, as well as the initial gas temperature in the ODT1, is very sensitive to the value of $r$.  During the rMOT cooling stage, 20-kHz drifts of the 741-nm laser frequency (linewidth 20 kHz) shift the cloud in $z$  enough to decrease transfer efficiency.  A Rb saturation-absorption reference laser stabilizes a transfer cavity to which the 741-nm laser is locked; together this provides sufficient frequency stability for BEC production.  

Following this procedure, we obtain up to $4$$\times$$10^{6}$ atoms in ODT1 at a temperature of 10 $\mu$K, which is 15$\times$ colder than the trap depth of ODT1.  We experimentally determine the polarizability of Dy at 1064 nm to be 116~a.u.~(atomic units) by measuring the ODT1 trap frequencies.  This value is 65\% smaller than recently calculated~\cite{Flambaum2011} and will be used to improve magic wavelength estimates for future Dy optical lattices.  

ODT1 is derived from a 30-W 1064-nm fiber laser, while ODT2 is derived from a 25-W diode-pumped Nd:YVO4 1064-nm laser.  Beams from both  are intensity-controlled by acousto-optic modulators (AOMs), and each beam is shaped into a cylindrical waist ($e^{-2}$ radius) focused at the cODT center.  The ODT1 has a waist of 30 (60) $\mu$m and trap frequency of 1~(0.5) kHz in $z$ ($\rho$), where $\rho$ lies in the $x$--$y$ plane. ODT2 has a waist of 22 (70) $\mu$m and trap frequency of 1.2~(0.4) kHz in $z$ ($\rho$).

After extinguishing the rMOT lasers and quadrupole field, we immediately spin-flip the atoms from the MWS-state to the maximally strong-field seeking (MSS) ground state to prevent heating from dipolar relaxation.  No optical pumping is needed because, as mentioned above, the rMOT---unlike a traditional MOT---provides a MWS-state spin-polarized gas~\cite{Berglund:2008}.  RF adiabatic rapid passage (ARP)  flips the spin within 2 ms from $m_{J}=8$ to $m_{J}=-8$ by using a $B_{x}=0.3$~G magnetic field and a swept-frequency RF-driven coil along $z$.  Stern-Gerlach measurements  indicate that  ARP produces a nearly pure, absolute ground state gas in $m_{J}=-8$, and we use such ARP sweeps and SG measurements to calibrate the applied magnetic fields and to zero the ambient magnetic field to $<$20 mG.  The 0.3 G $B_{x}$ field ($\Delta m = 1$ Zeeman energy equal to 25 $\mu$K) is then rotated to point along $z$ and increased to $B_{z}=2.3$ G (190 $\mu$K) to suppress thermally driven spin transfer to $m_{J}>-8$ states. 

\begin{figure}[t]
\includegraphics[width=0.4\textwidth]{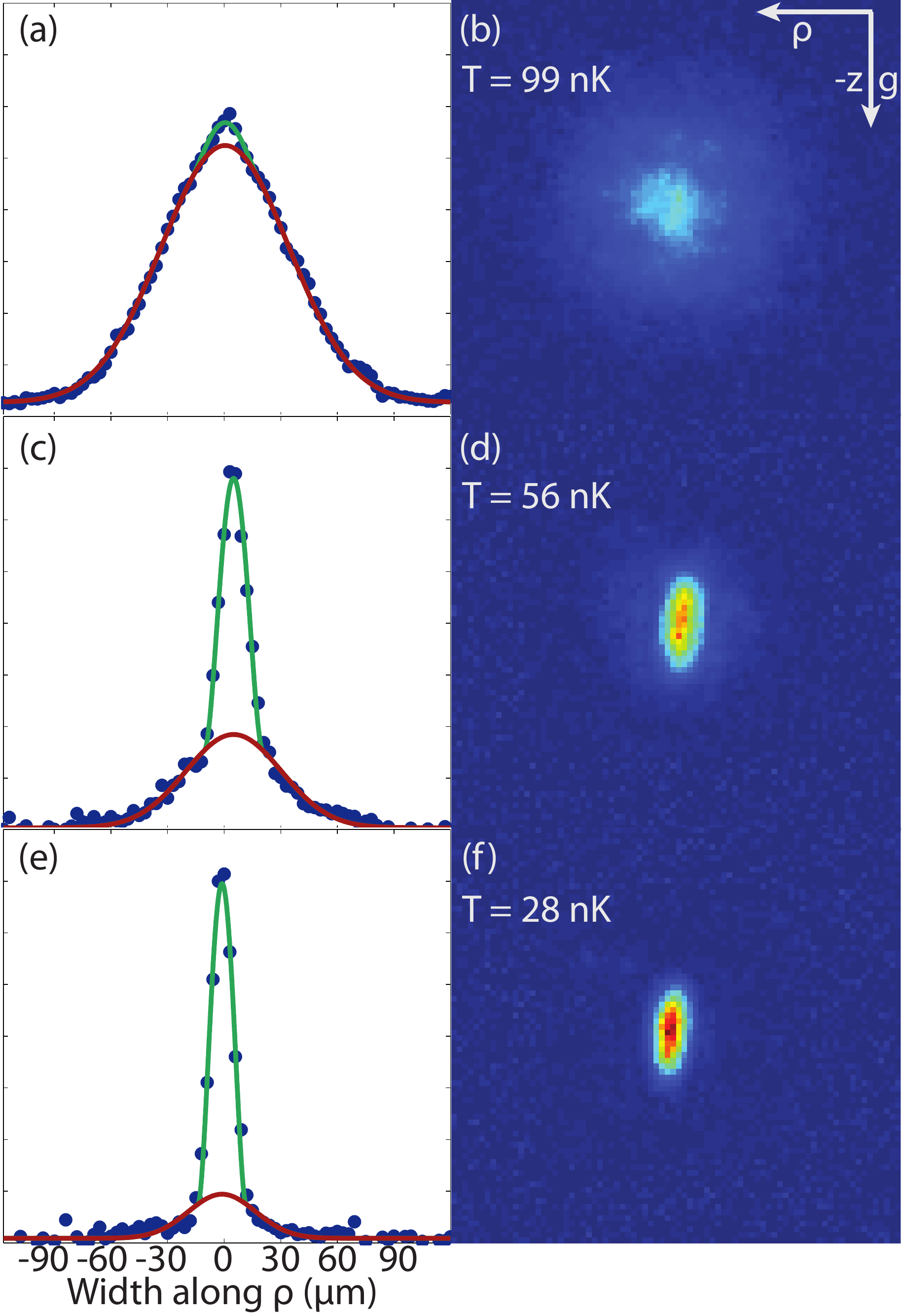}
\caption{(color online). TOF profiles of the spin-purified Dy gas for three evaporation time-constants, with $\tau=15$ s in (e) and (f).  (a,c,e) Data at centers are fit to a parabolic profile (green), which underestimates condensate fraction, whereas the distributions' wings are fit to a Gaussian profile (red).  (b,d,f) Absorption images of the emerging BEC.  (b) The transition temperature is 99(5) nK, with condensate fraction 14.4(3)\%; (b) 44(2)\% condensate fraction at 56(3) nK; (c) a BEC of condensate fraction 73(4)\% and 1.5(2)$\times$$10^{4}$ atoms forms at 28(2) nK with density 10$^{14}$ cm$^{-3}$.}
\label{BEC}
\end{figure}

Within 4 s of loading Dy from  ODT1 into the 340~$\mu$K  cODT, formed by ramping  ODT2 to 12~W in 1 s, plain evaporation both reduces the atom number to 1$\times$$10^{5}$ at the cODT center and decreases the temperature to 2 $\mu$K.  The trap frequencies are $[f_{x},f_{y},f_{z}]=[380, 500, 1570]$ Hz before the forced evaporation sequence begins.  Power in the ODTs are reduced using the functional form $P_{f}=P_{i}/(1+t/\tau)^{\beta}$~\cite{Thomas01} using experimentally optimized parameters $\beta_{1}=2$ and $\tau=15$ s for ODT1.  After a 9 s  delay, ODT2's power is reduced for 23 s using  $\beta_{2}=1.3$ and $\tau=15$ s.  To recompress the trap and obtain an oblate shape---$f_{z}/f_{\rho}>1$,  with $f_{\rho}=\sqrt{f_{x}f_{y}}$---both beams are linearly ramped in 0.8 s to their final powers.  For the data presented here,  final ODT powers yield the trap frequencies  $[f_{x},f_{y},f_{z}]=[205, 195, 760]$~Hz, which provide a ratio $f_{z}/f_{\rho} = 3.8$~\footnote{Trap frequencies, measured via parametric heating, are uncertain to  5\%.}.  The BEC is held for 0.2 s, the first 0.1~s of which the $B_{z}$ field may be rotated into the $\rho$-plane.  Time-of-flight measurements are made by rapidly extinguishing the cODT and, after 1~ms, rotating the field into the imaging axis [see Fig.~\ref{Scheme}(b)].  Absorption images are obtained using resonant 421-nm light and an exposure of 60 $\mu$s, much less than the decay time to metastable states~\cite{Lu2010,*Youn2010a}.  

The emergence of the Dy BEC below $T=100$ nK is shown in Fig.~\ref{BEC}.  A dense parabolic peak emerges from a Gaussian background signaling the formation of a condensate with high purity below $T=30$ nK.   The DDI preserves a parabolic condensate profile~\cite{Santos03}, but the criterion for the Thomas-Fermi (TF) limit is modified for a dipolar BEC.  Approximate expressions for the dipolar TF limit in the spherical and highly elongated prolate and oblate BEC regimes have been obtained~\cite{ODell08}.  Although our moderately oblate BEC fails to fall into these regimes, we may use the oblate BEC analysis in Ref.~\cite{ODell08} to estimate that the Dy BEC satisfies the dipolar TF limit for ``effective'' $a_{s}$'s~\footnote{$^{164}$Dy has 153 Born-Oppenheimer potentials~\cite{Kotochigova11} which can contribute to the effective $a_{s}$.} in the vicinity of the recently estimated van der Waals length $\bar{a}=76a_{0}$~\cite{Kotochigova11}.   

\begin{figure}[t!]
\includegraphics[width=0.4\textwidth]{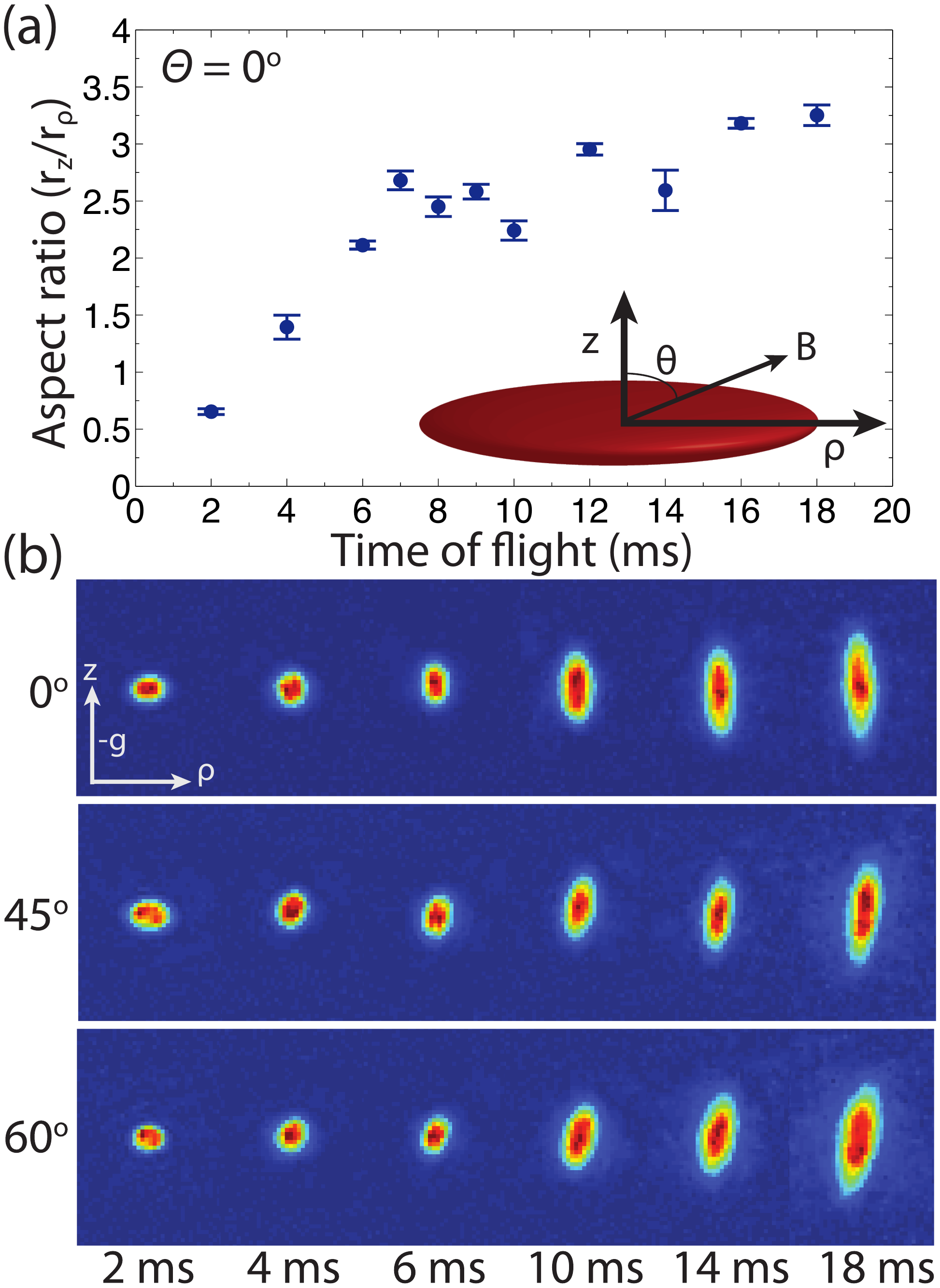}
\caption{(color online). Anisotropic expansion profile versus time after trap release.  (a) $r_{z}$ and $r_{\rho}$ are the dimensions of the parabolic profile fit to the BEC for $\theta=0^{\circ}$.  Inset:  Schematic of the oblate trap and magnetic field orientation.  (b) Images of the expanding condensate after trap release.  The condensate rotates by 7(1)$^{\circ}$ [9.4(6)$^{\circ}$] with respect to the $\theta=0^{\circ}$ expansion orientation for  $\theta=45^{\circ}$ [$\theta=60^{\circ}$].  No BEC forms for $\theta=90^{\circ}$.}
\label{expansion}
\end{figure}

We calculate that an initial phase space density for BEC is reached at 120(20) nK for a trap with our mean frequency 310 Hz and atom number 7$\times$$10^{4}$.  Applying corrections~\cite{Pfau05CrBEC} for finite size effects and mean field energy (using $a_{s} = 100a_{0}$), we estimate a $T_{c}=100(20)$~nK, which is consistent with our measurement of 99(5) nK.  The nearly pure BEC in panels (e) and (f) contains 1.5$\times$$10^{4}$ atoms at a density of 10$^{14}$ cm$^{-3}$.   Further evidence of $^{164}$Dy's Bose-condensation is provided in Fig.~\ref{expansion}, which shows the characteristic aspect ratio inversion of an interacting BEC released from an anisotropic trap.  These data are obtained with a small, $B_{z}=2.3$ G  field parallel to the oblate trap's axis along $z$.  BECs also form with 2.3 G fields rotated to an angle either $\theta=45^{\circ}$ or  60$^{\circ}$ with respect to the $z$-axis.  However, no BEC forms in a 90$^{\circ}$ field ($B_{z}=0$, $B_{\rho}=2.3$ G), and a thermal cloud is  obtained regardless of evaporation  parameters.  

Strongly dipolar BECs---defined as $\epsilon_{dd}>1$---have been shown~\cite{Santos00,*You01,Eberlein04,*Eberlein05,Pfau08Collapse,PfauReview09} to be unstable for harmonic traps with $\gamma=f_{\parallel}/f_{\perp} <1$ but metastable for traps with $\gamma=f_{\parallel}/f_{\perp} >1$, where $f_{\parallel}$ is the trap frequency parallel to the polarizing field, and $f_{\perp}$ is the geometric mean of the trap frequencies perpendicular to the field. The attractive part of the DDI acts to elongate the cloud along the field direction, which if pointed along the trap's major axis leads to an instability much akin to that of non-dipolar BECs with large negative $a_{s}$~\cite{PfauReview09}.  In contrast, aligning the field along the minor axis acts to suppress this elongation, minimizing the attractive contribution from the DDI to the mean field energy and stabilizing the gas.  Weakly dipolar BECs---wherein $\epsilon_{dd}<1$---are stable regardless of the value of $\gamma$, because the repulsive contact interaction  dominates the DDI~\cite{PfauReview09}.

While we observe a BEC in our oblate trap for $\theta=0^{\circ}$ ($\gamma = 3.8$), the failure to observe a BEC in the oblate trap with $\theta = 90^{\circ}$ ($\gamma=0.51$) is strongly suggestive of a BEC in the strongly dipolar regime~\footnote{For the field orientation $\theta = 0^{\circ}$ ($90^{\circ}$), $\gamma$ is defined using $f_{\parallel}=f_{z}$ ($f_{\parallel}=f_{\rho}$) and $f_{\perp}=f_{\rho}$ ($f_{\perp}=\sqrt{f_{\rho}f_{z}}$).  Most references analyze BEC stability for traps cylindrically symmetric about the polarization axis ($\theta=0^{\circ}$).  However, an oblate trap with our aspect ratio and $\theta=90^{\circ}$ was treated in Ref.~\cite{LimaThesis}  with the result that $\epsilon_{dd}=1$ remains the instability boundary.}.  Moreover, the tilt of the expanding condensate toward the polarizing field is indicative of a strongly dipolar condensate and was not observed in the weakly dipolar Cr system~\footnote{T.~Pfau, private communication (2011).}.  Gross--Pitaevskii equation simulations of the BEC stability boundary and tilt angle versus $\theta$, along with simulations of the TOF aspect ratio evolution, should provide a first measurement of $^{164}$Dy's effective $a_{s}$~\footnote{R.~Wilson and J.~Bohn, private communication (2011).}.

%Simulations using our trap parameters of the BEC stability boundary versus $\theta$, along with simulations of the TOF aspect ratio evolution, should provide a first measurement of $^{164}$Dy's effective $a_{s}$.  An initial attempt failed to fit the TOF data in Fig.~\ref{expansion}(a) when using the formalism of Ref.~\cite{Giovanazzi06}, which assumes the TF regime and a separable wavefunction.  However, this latter assumption is less well-founded in the $\epsilon_{dd}>1$ regime, as recently confirmed in Gross--Pitaevskii equation simulations of our system parameters~\footnote{R.~Wilson and J.~Bohn, private communication (2011).}.  A forthcoming publication will present  these simulation results, but using the condition imposed by $\epsilon_{dd}>1$, we can presently place the  first bound  $a_{s}\alt130a_{0}$ on $^{164}$Dy's effective scattering length.

In future work, we will search for broad Feshbach resonances and load the Dy BEC into optical lattices to search for phases predicted by the extended Bose-Hubbard model~\cite{PfauReview09}.  Quantum degenerate gases of Dy's other isotopes, including dipolar Bose-Fermi mixtures, could be formed with the experimental technique presented here.  Indeed, we recently optically dipole-trapped an ultracold dipolar Bose-Fermi mixture of $^{164}$Dy and $^{163}$Dy---the first of its kind---using a modified version of this experimental technique; ongoing work strives to cool the mixture to quantum degeneracy.

We thank R. Wilson, J. Bohn, and S. Kotochigova for helpful discussions.  We acknowledge support from the NSF, AFOSR, ARO-MURI on Quantum Circuits, and the David and Lucille Packard Foundation.

%\bibliography{Lev_Refs}

\end{document}